\documentclass[11pt,a4paper]{article}

\usepackage{jheppub}

\usepackage{amsmath}
\usepackage{amscd}
\usepackage{amsfonts}
\usepackage{amssymb}
\usepackage{mathrsfs}
\usepackage{fancybox} 
\usepackage{enumerate}

\usepackage{graphicx}
\usepackage{url} 

\usepackage{color}
\usepackage{ulem}

\usepackage{comment}

\def\={\stackrel{\bullet}{=}}

\def\({\left(}
\def\){\right)}
\def\[{\left[}
\def\]{\right]}

\def \be {\begin{equation}}
\def \ee {\end{equation}}
\def \beqa {\begin{eqnarray}}
\def \eeqa {\end{eqnarray}}
\def \beal#1 {\begin{align}#1\end{align}}
\def \bes#1 {\begin{equation}\begin{split}#1\end{split}\end{equation}}
\def \nn {\notag\\}

\def\ue#1#2{\overset{#1}{#2}}

\def\xt{(\boldsymbol{x},t)}
\def\x't{(\boldsymbol{x'},t)}
\def\xnt{(\boldsymbol{x_n},t)}
\def\vx{\boldsymbol{x}} 
\def\vxn{\boldsymbol{x}_n} 
\def\vxk{\boldsymbol{x}_k} 
\def\vv{\boldsymbol{v}} 
\def\vvn{\boldsymbol{v}_n} 
\def\vr{\boldsymbol{r}} 
\def\vV{\boldsymbol{V}} 

\def\abs#1#2#3{\vert #1#2#3 \vert} 
\def\chris#1#2#3{\Gamma^{#1}_{\,#2#3}}

\def\3tensor#1#2#3#4{#1^{#2\;#4}_{\;\;#3}}
\def\rg{\sqrt{-g} } 
\def\ADM#1{#1_{\text{ADM}}} 

\begin{document}

\vspace{-2cm}
\begin{flushright}
\normalsize{ 
YITP-23-66
\\ OU-HET-1188}
\end{flushright}

\title{Energies and a gravitational charge for massive particles in general relativity }
\author[a]{ Sinya Aoki\footnote{\it saoki@yukawa.kyoto-u.ac.jp },}
\affiliation[a]{Center for Gravitational Physics, Yukawa Institute for Theoretical Physics, Kyoto University, Kitashirakawa Oiwakecho, Sakyo-ku, Kyoto 606-8502, Japan}
\author[a,b]{ Tetsuya Onogi\footnote{\it onogi@phy.sci.osaka-u.ac.jp},}
\affiliation[b]{Department of Physics, Osaka University, Toyonaka, Osaka 560-0043, Japan}
\author[b]{Tatsuya Yamaoka\footnote{\it t\_yamaoka@het.phys.sci.osaka-u.ac.jp}\,}

\abstract{
In this paper, we investigate relations or differences among various conserved quantities which involve the matter Energy Momentum Tensor (EMT) in general relativity. These quantities include 
the energy with Einstein's pseudo EMT, the generalized Komar integral, or the ADM energy,  all of which can be derived from Noether's second theorem, as well as an extra conserved charge recently proposed in general relativity.
For detailed analyses,  we apply definitions of these charges to a system of free massive particles. We employ the post-Newtonian (PN) expansion to make physical interpretations.
We find that the generalized Komar integral is not conserved at the first non-trivial order in the PN expansion due to non-zero contributions at spatial boundaries,
while the energy with  Einstein's pseudo EMT 
at this order agrees with a total energy of massive particles with gravitational interactions through the Newtonian potential,
 and thus is conserved.  In addition, this total energy is shown to be identical to the ADM energy not only at this order but also
 all orders in the PN expansion.
We next calculate an extra 
conserved charge for the system of massive particles, at all orders in the PN expansion,
which turns out to be a total number of particles.
We call it a gravitational charge, since it is clearly different from the total energy. 
We finally discuss an implication from a fact that there exist two conserved  quantities, energy and gravitational charge, in general relativity. 
}

\date{today}

\maketitle



\section{Introduction}
\label{intro}
Even though general relativity was constructed on the basis of two fundamental assumptions,
the general covariance and the equivalent principle\cite{Einstein:1916vd},
it has been non-trivial to establish a covariant definition of a conserved energy.
Einstein turned the covariant conservation law of the energy momentum tensor (EMT) as $\nabla_\mu T^{\mu\nu}=0$ to
a conservation law in a curved spacetime as $\partial_\mu \{ \sqrt{-g} (T^{\mu\nu} +\tilde t^{\mu\nu} )\} =0$
by introducing non-covariant pseudo-tensor $\tilde t^{\mu\nu}$.  (For example, see Ref.~\cite{Misner:1973prb}.)
He then defined a total energy  as a volume integral of $T^{\mu\nu} +\tilde t^{\mu\nu}$, and interpreted that
$\tilde t^{\mu\nu}$ represents an energy of the gravitational field. 
As $\sqrt{-g} (T^{\mu\nu} +\tilde t^{\mu\nu} )$ is expressed in term of  a total derivative of another pseudo-tensor,
the total energy is given by a surface integral via Stokes' theorem, which is called a quasi-local (expression of) energy.
Later there appeared many different definitions for the quasi-local energy, which include the Komar energy\cite{Komar:1958wp} and the ADM energy\cite{Arnowitt:1962hi}. 
Recently it is pointed out\cite{Aoki:2022gez} that the conservation of  quasi-local energies including the pseudo-tensor can be understood  from 
Noether's second theorem\cite{Noether:1918zz}.

Under these circumstances, one may ask whether all quasi-local energies are indeed conserved for a given system.
Since it is unlikely that there exists many different conserved energies, possibilities are
either different definitions give a same conserved energy or only some of them are conserved but others are not.

Recently two of the present authors with their colleague have made two proposals\cite{Aoki:2020prb,Aoki:2020nzm}
that (1) a matter energy can be covariantly defined in general relativity from the EMT though it is
conserved only for special cases and not conserved in general, and (2) it is always possible to construct an extra conserved charge from the EMT, which is different from the matter energy. 
While the extra conserved charge can be interpreted as an entropy for special cases such as the FLRW expanding universe\cite{Aoki:2020nzm,Aoki:2022gez}, 
more investigations are required to give a definite interpretation for general cases.

In this paper,
in order to answer the questions raised above, we study the properties of 
various conserved quantities which involve the matter EMT, by applying them to a system with massive point particles interacting only through the gravitational field in general relativity.
For  the first question, we focus on the matter energy as well as (quasi-local) energies with two different definitions of pseudo-tensors, one is Einstein's pseudo-tensor, the other is related to it by a total divergence.
We calculate three energies for this simple  system at the first non-trivial order in the post-Newtonian expansion (1PN).
Results are summarized as follows. 
The energy with Einstein's pseudo-tensor is shown to be conserved at the 1PN,
while the matter energy and the energy with the other pseudo-tensor are not conserved.
Even though the energy with the other pseudo-tensor  is constructed from the conserved current,
we show that flows at spacial boundaries cause its non-conservation.
We also argue that the energy with Einstein's pseudo-tensor agrees with the ADM energy at all orders in the PN expansion
for an asymptotically flat spacetime.
For the second question, we evaluate the extra conserved charge for this system and show that it gives 
a total number of the massive particles 
in the system at all orders in the PN expansion.

The paper is organized as follows.
In Sec.~\ref{sec:energy}, we introduce three definitions of energies. 
In particular, we show that different pseudo-tensors can be easily constructed by Noether's second theorem.
In addition, two pseudo-tensors used in this paper are presented.
 A construction as well as some issues of the extra conserved charge are given in Sec.~\ref{sec:extra}.
In Sec.~\ref{sec:system}, we explain our setup for a system with massive point particles in general relativity,
and we present  formulae in the PN expansion necessary in our calculation,
which are mainly taken from Ref.~\cite{Weinberg:1972kfs}.
Our main results are given in \ref{sec:res_energy} and \ref{sec:res_charge}.
Some results used in the main text are derived in appendices~\ref{app:PN} and \ref{app:Integration}.

\section{Energies and  an extra conserved charge}
\label{sec:definition}
We here give three different definitions of energies and an extra conserved charge different from energies in general relativity,
all of which will be used in this paper.

\subsection{Different definitions of energy}
\label{sec:energy}
\subsubsection{A matter energy}
The first definition of energy we consider in general relativity is a matter energy\cite{Aoki:2020prb,Aoki:2022gez}, given by
\beqa
E_M(x^0) &:=& \int_{x^0} [d^3 x]_\mu T^\mu{}_\nu \tau^\nu = - \int_{x^0} [d^3 x]_\mu T^\mu{}_0, \quad \tau^\mu := -\delta^\mu_0,
\label{energy_matter}
\eeqa  
where $T^\mu{}_\nu$ is an EMT appearing in Einstein equation as $G^{\mu\nu} = 2\kappa T^{\mu\nu}$ with $\kappa := 4\pi G$, and
an integral is performed on the constant $x^0$ hyper-surface with a volume element   $ [d^3 x]_\mu$.
We here choose a vector $\tau^\mu$ to define an energy, since $\tau^\mu$ corresponds to a generator for a global time translation of the system. Thus 
this energy is not conserved, {\it i.e.} $\dfrac{d E(x^0)}{dx^0}\not= 0$, in general due to a violation of the global time translation of a matter action,
except for special cases\cite{Aoki:2022gez}.
For example, if $\tau^\mu$ is a Killing vector of the system, $T^\mu{}_\nu \tau^\nu$ becomes a conserved current, and thus $E_M(x^0)$ is conserved. 

\subsubsection{Energies from a pseudo-tensor}
In the second and third definitions, energy is defined from a pseudo-tensor as
\beqa
E_{\rm pt}(x^0) &=& - \int_{x^0} [d^3 x]_\mu ( \frac{1}{2\kappa}G^\mu{}_0 + t^\mu{}_0),
\label{energy_ps}
\eeqa
where $t^\mu{}_\nu$ is called a pseudo-tensor, since it is not covariant under the general coordinate transformation. 
There are several important properties for this definition.
First of all, it satisfies
\beqa
\partial_\mu\[ \sqrt{-g} \( \frac{1}{2\kappa}G^\mu{}_\nu + t^\mu{}_\nu \)\] &=&0
\label{conv_ps}
\eeqa
as an identities without using equations of motion (EOM), where $g$ is a determinant of the metric $g_{\mu\nu}$.
Note that this condition is not covariant, so that $t^\mu{}_\nu$ is also not covariant.
It is also noted that \eqref{conv_ps}  follows immediately  from \eqref{quasi_ps}.
Secondly, it can always be written as a total derivative of an anti-symmetric quantity,
\beqa
\sqrt{-g} \( \frac{1}{2\kappa}G^\mu{}_\nu+ t^\mu{}_\nu \) = \partial_\rho X^\rho{}_\nu{}^\mu,
\label{quasi_ps}
\eeqa
where $X^\rho{}_\nu{}^\mu$ is anti-symmetric under an exchange of $\rho$ and $\mu$ as $X^\mu{}_\nu{}^\rho=-X^\rho{}_\nu{}^\mu$,
so that $E_{\rm pt}(x^0)$ can be written as a surface integral via Stokes' theorem.
If we use the Einstein equation, it can also be written as
\beqa
E_{\rm pt}(x^0) &\underset{\rm EOM}{=}& - \int_{x^0} [d^3 x]_\mu ( T^\mu{}_0 + t^\mu{}_0),
\label{energy_ps2}
\eeqa
from which one interprets the contribution from $t^\mu{}_0$ as the gravitational energy.

These two properties are consequences of Noether's 2nd theorem\cite{Aoki:2022gez}. 
In the case of Einstein-Hilbert action that
\beqa
S_G = {1\over 4\kappa} \int d^4x\, \sqrt{-g} ( R- 2\Lambda),
\eeqa
where $R$ is the Ricci scalar and $\Lambda$ is the cosmological constant, the theorem implies an existence of a conservation current density,
which is explicitly given by
\beqa
\sqrt{-g}J^\mu[\xi] &:=& {1\over 4\kappa} \sqrt{-g} \nabla_\nu \left[ \nabla^{[\mu} \xi^{\nu]}\right]
= A^\mu{}_\nu \xi^\nu + B^\mu{}_\nu{}^\rho \xi^\nu_{,\rho} + C^\mu{}_\nu{}^{\rho\lambda} \xi^\nu_{,\rho\lambda},
\label{def_J}
\eeqa
where an arbitrary vector $\xi^\nu$ corresponds to a generator of the general coordinate transformation,
$ f^{[\mu,\nu]} := f^{\mu\nu} - f^{\nu\mu}$ defines an anti-symmetrization of indices and 
$f_{,\mu} :=\partial_\mu f$ denotes a partial derivative.
As seen from the first expression in the above, it is easy to show  $\partial_\mu(\sqrt{-g}  J^\mu[ \xi]) =0$ for an arbitrary vector $\xi$ 
without EOM. 
The conservation $\partial_\mu (\sqrt{-g}J^\mu[ \xi]) =0$ applied to the second expression also implies\cite{Aoki:2022gez} 
\beqa
\partial_\mu A^\mu{}_\nu &=&0, \quad  A^\mu{}_\nu = -\partial_\rho \tilde B^\rho{}_\nu{}^\mu,
\quad \tilde B^\rho{}_\nu{}^\mu := {1\over 2} B^{[\rho}{}_\nu{}^{\mu]}-{1\over 3} \partial_\lambda C^{[\rho}{}_\nu{}^{\mu]\lambda},
\eeqa
which correspond to two properties, \eqref{conv_ps} and \eqref{quasi_ps},
if we identify $A^\mu{}_\nu = \sqrt{-g}(\frac{1}{2\kappa}G^\mu{}_\nu+ t^\mu{}_\nu)$.
It is noted that one may also construct a $\xi$-dependent conserved charge, called a generalized Komar energy\cite{Komar:1958wp}, from the conserved current $J^\mu[\xi]$ as\cite{Aoki:2022gez}
\beqa
E_{\rm Komar} [\xi] &=& \int_{x^0} [d^3 x]_\mu J^\mu[\xi].
\label{Komar}
\eeqa
One may regard the energy from the pseudo-tensor  in \eqref{energy_ps} as a generalized Komar energy for a particular choice of the vector that $\xi^\mu =\tau^\mu$.
An explicit form of $t^\mu{}_\nu$ is given by\cite{Aoki:2022gez} 
\beqa
t^\mu{}_\nu &=&{1\over 4\kappa}\left[ (R-\Lambda)\delta^\mu_\nu + g^{\rho\mu} \Gamma^\lambda_{\lambda\nu,\rho} -
 g^{\rho\lambda} \Gamma^\mu_{\lambda\rho,\nu}\right],
 \label{EH_pt}
\eeqa
which contains second derivatives of the metric $g_{\mu\nu}$.
It is clear that $t^\mu{}_\nu$ is not a covariant tensor, since it contains partial derivatives of $\chris\mu\nu\alpha$. 
We call $t^\mu{}_\nu$  in \eqref{EH_pt} an Einstein-Hilbert (EH) pseudo-tensor, and
$E_{\rm pt}(x^0)$ with the EH  pseudo-tensor is referred to an EH pseudo-tensor energy for short.

Without changing the Einstein equation, it is allowed to add
a total derivative term to  the Einstein-Hilbert action as
\beqa
S_G\to S_G -{b\over 4\kappa}\int d^4x  \partial_\mu D^\mu
\eeqa
where $b$ is a real parameter and
\beqa
{\cal D}^\mu &:=&\sqrt{-g} \left[ g^{\mu \alpha} \Gamma^\beta_{\alpha\beta} -g^{\alpha\beta}\Gamma^\mu_{\alpha\beta}\right].
\eeqa
Consequently, the corresponding pseudo-tensor becomes
\beqa
\sqrt{-g}t^\mu{}_\nu [b] =\sqrt{-g} t^\mu{}_\nu + {b\over 4\kappa} \partial_\alpha \left( \delta^{[\mu}_\nu D^{\alpha]}\right),
\eeqa
which satisfies $\partial_\mu \{\sqrt{-g}(T^\mu{}_\nu+t^\mu{}_\nu [b])\}=0$ since the second term trivially vanishes.
In particular at $b=1$, we have  
\beqa
\tilde t^\mu{}_\nu &:=& t^\mu{}_\nu [b=1]={1\over 4\kappa}\left[\delta^\mu_\nu g^{\alpha\beta} \left( \chris\rho\alpha\lambda \chris\lambda\beta\rho
- \chris\rho\alpha\beta \chris\lambda\beta\lambda \right)
\right],
\eeqa
which consists of the metric $g_{\mu\nu}$ and its first derivatives without its second derivatives\cite{Utiyama:1984bc}, and
we define the corresponding energy as 
\beqa
\tilde E_{\rm pt}(x^0) &=& - \int_{x^0} [d^3 x]_\mu \( \frac{1}{2\kappa}G^\mu{}_0 + \tilde t^\mu{}_0\) 
\underset{\rm EOM}{=}- \int_{x^0} [d^3 x]_\mu ( T^\mu{}_0 + \tilde t^\mu{}_0).
\label{energy_ps2}
\eeqa
We call $\tilde t^\mu{}_\nu$ and $\tilde E_{\rm pt}(x^0)$ an Einstein's pseudo-tensor and an Einstein's pseudo-tensor energy, respectively.
 
One may wonder whether $E_{\rm pt}(x^0)$ and $\tilde E_{\rm pt}(x^0)$ are both conserved, since
it seems strange that there exist two definitions of the conserved energy.
In this paper, we will calculate $E_{\rm pt}(x^0)$ and $\tilde E_{\rm pt}(x^0)$ as well as $E_M(x^0)$
for massive point particles in general relativity at the 1PN,
in order to determine which energy is conserved.

\subsection{An extra conserved charge}
\label{sec:extra}
In Ref.~\cite{Aoki:2020nzm},
it has been pointed out that there alway exists an extra conserved charge in general relativity, which
seems different from energies, $E_M(x^0)$,  $E_{\rm pt}(x^0)$ or $\tilde E_{\rm pt}(x^0)$. 

We first  construct a conserved current $K^\mu$ form the EMT as $K^\mu := T^\mu{}_\nu \beta^\nu$,
whose conservation leads to a equation for $\beta^\nu$ as
\beqa
\nabla_\mu K^\mu = \nabla_\mu (T^\mu{}_\nu \beta^\nu) = T^\mu{}_\nu \nabla_\mu \beta^\nu = 0,
\label{cond_zeta}
\eeqa
where we use $\nabla_\mu T^\mu{}_\nu =0$ and summations over $\mu,\nu$ are taken.
The corresponding conserved charge is defined by
\beqa
Q(x^0) &=& \int_{x^0} [d^3x]_\mu T^\mu{}_\nu \beta^\nu ,
\eeqa
which looks a generalization of $E_M(x^0)$ in \eqref{energy_matter}.
This charge is understood in terms of the Noether's first theorem for the matter action\cite{Aoki:2022ugd}.

There are two main issues for this conserved charge.
First of all, we would like to understand a physical meaning of the conserved charge if it is indeed different from energy.
It is argued that the conserved charge may be regarded as an entropy for special cases\cite{Aoki:2020nzm, Aoki:2022gez,Aoki:2022tek},
but it may be more general than the entropy as suggested in \cite{Aoki:2022ysm}.
In this paper we calculate the conserved charge $Q(x^0)$ for a simple system, non-interacting massive point particles in general relativity,
in order to have a hint for its physical interpretation.
Secondly, we should provide a definite and general principle to choose a direction of $\beta^\mu$, which is necessary to
determine it  from a scalar equation \eqref{cond_zeta}.
In previous studies\cite{Aoki:2020nzm, Aoki:2022gez,Aoki:2022tek}, its direction was taken case by case, 
but a definite principle for a choice seems necessary for its correct physical interpretation. 
Recently it has been proposed that the direction of $\beta^\mu$ is determined from the EMT\cite{Aoki:2022ysm}. 
In this paper, we test this proposal for massive point particles in general relativity.

\section{A system of massive point particles in general relativity}
\label{sec:system}
\subsection{Setup}
\label{sec:setup}
In this paper, we consider a system of massive point particles interacting only through gravitational interaction.
An action of the system is given by $S = S_G + S_M$ with 
 \begin{gather}
       S_M = -\sum^N_{n=1} \frac{1}{2} \int ds_n \[m_n^2 e_n (s_n)- \frac{g_{\mu\nu}}{e_n(s_n)} v^\mu_n v^\nu_n \],
       \label{SM}
 \end{gather}
 where  $N$ is a number of massive particles, $m_n, s_n$ and $e_n(s_n)$ are mass,  proper time and einbein  of the $n$-th particle, 
 respectively. A spacetime position and a 4 velocity of the $n$-th particle are denoted by $x_n:= x_n(s_n)$ and $v_n(s_n) := \dfrac{dx_n(s_n)}{ds_n}$.
 Note that $S_M$ is invariant under the reparametrization of each proper time $s_n\to s_n^\prime$ followed with
\begin{equation}
       e_n(s_n) \rightarrow e'_n(s'_n) := e_n(s_n)\frac{ds_n}{ds'_n}.
\end{equation}

Equations of motion derived from the above action lead to
\beqa
       G^{\mu\nu} + \Lambda g^{\mu\nu} &=& 2\kappa T^{\mu\nu},\\
       e_n (s_n) &=& \frac{1}{m_n}\sqrt{-g_{\alpha\beta}(x_n)v_n^\alpha v_n^\beta},
       \label{einbein}\\
     {d\over d s_n} v_n^\mu (s_n) + \chris\mu\alpha\beta v_n^\alpha(s_n)v_n^\beta(s_n) &=&  {d\over d s_n}\log e_n(s_n), \quad n=1,2,\cdots,N,
         \label{geodesic} 
\eeqa
where an EMT for massive particles is given by
\begin{equation}
       \label{EMT}
       T^{\mu\nu}:= \frac{2}{\sqrt{-g}}\frac{\delta S_M}{\delta g_{\mu\nu}(x)} ={1\over \sqrt{-g}}  \sum^N_{n=1} \int \frac{ds_n}{e_n(s_n)} v_n^\mu(s_n) v_n^\nu (s_n)\delta^{(4)}(x - x_n(s_n)),
\end{equation}
which satisfies $\nabla_\mu T^{\mu\nu}=0$.

Using the reparametrization invariance, we take the affine parameter $s_n$ for each $n$ to satisfy $m_ne_n(s_n)=1$, 
so that \eqref{einbein} and \eqref{geodesic} of EOM become
 \beqa
g_{\mu\nu}(x_n)v_n^\mu(s_n)v_n^\nu(s_n) &=&-1  
\label{normal}\\
   {d \over ds_n} v_n^\mu(s_n)+\chris\mu\alpha\beta v_n^\alpha (s_n)v^\beta_n(s_n)&=&0, \quad n=1,2,\cdots,N,
   \label{geodesic2}
 \eeqa
 where
\beqa
       T^{\mu\nu} &=& \frac{1}{\sqrt{-g}} \sum^N_{n=1}m_n \int ds_n\,v_n^\mu(s_n)v_n^\nu(s_n)\delta^{(4)}(x -x_n(s_n)).
\eeqa
If $v_n^\mu(s_n)$ satisfies \eqref{geodesic2}, the normalization condition \eqref{normal} for $v_n^\mu(s_n)$ does not depend on $s_n$, so that
we need to only require \eqref{normal} as an initial condition at some $s_n$.   

Hereafter, we set $\Lambda=0$ otherwise stated.

\subsection{Post-Newtonian approximation}
\label{sec:post-Newtonian}
We analyze the system in the post-Newtonian approximation by expanding  motions of matter and the metric field in powers of $v/c$. 
In the following, 
we will use results in Chapter 9 of the Weinberg's  textbook\cite{Weinberg:1972kfs},
which are summarized below.

A metric tensor is expanded as
\beqa
       g_{00} &=& -1 + \ue2{g_{00}} + \ue4g{_{00}}+ \cdots, \
       g_{ij} = \delta_{ij} + \ue2{g_{ij}} + \cdots, \ 
       g_{i0} = \ue3{g_{i0}} +\cdots, 
\eeqa
where
\beqa
       \ue2{g_{00}}=-2\phi, \quad \ue4{g_{00}}=-2\phi^2-2\psi, \quad\ue2{g_{ij}}= -2\delta_{ij}\phi, \quad
       \ue3{g_{i0}}=\zeta_i,\
       \label{Ngab}
\eeqa
and 
\beqa
       \phi \xt &=& -G\int d^3 x' \frac{\ue0{T^{00}}\x't }{\vert \vx - \vx' \vert}, \quad
       \zeta_i \xt = -4G \int d^3x' \frac{\ue1{T^{i0}}\x't}{\vert \vx - \vx' \vert}, \\
       \psi \xt &=& -\int {d^3 x'\over \vert \vx - \vx' \vert}\left[ {1\over 4\pi}  \phi_{,00}\x't
       + G \ue0{T^{00}}\x't +  G \ue2{T^{ij}}  \x't \right] 
\eeqa
with $t:=x^0$. Then  the determinant of $g_{\mu\nu}$ is given by $g= -1 +4\phi+\cdots$.

The EMT appeared in the above equations is expanded as
\beqa
       \ue0{T^{00}} &=& \sum_n m_n \delta^3(\vx -\vxn), \quad 
       \ue2{T^{00}} = \sum_n m_n(\phi + \frac{1}{2}\vvn ^2 )\delta^3(\vx - \vxn),\ \vvn^2:=\sum_{i=1}^3 (v_n^i)^2,
       \nn
       \ue1{T^{i0}} &=& \sum_n m_nv_n^i \delta^3(\vx - \vxn),\quad
       \ue2{T^{ij}} = \sum_n m_n v_n^i v_n^j \delta^3(\vx - \vxn).
       \label{2Tij}
\eeqa
Therefore we have
\beqa
 \phi \xt &=& -G \sum_n {m_n \over \vert \vx - \vxn \vert}, \quad
  \zeta_i \xt = -4G \sum_n {m_n v_n^i\over \vert \vx - \vxn \vert}.
  \label{phi-zeta}
\eeqa

In this paper, we also use the following results.
\beqa
\ue2{\chris{i}{0}{0}} &=& \phi_{,i}, 
\quad  
\ue4{\chris{i}{0}{0}} =\( 2\phi^2+\psi\)_{,i} +\partial_t \zeta_{i,0},
\quad
\ue3{\chris{i}{0}{j}} ={1\over 2}\( \zeta_{i,j} -\zeta_{j,i}\) -\delta_{ij} \phi_{,0},\nn
\ue2{\chris{i}{j}{k}} &=&-\delta_{ij} \phi_{,k} -\delta_{ik} \phi_{,j} +\delta_{jk}\phi_{,i}, 
\quad  
\ue3{\chris{0}{0}{0}} = \phi_{,0},
\quad
\ue2{\chris{0}{0}{i}} = \phi_{,i},\nn
\ue3{\chris{0}{i}{j}} &=&-{1\over 2}\( \zeta_{i,j} +\zeta_{j,i}\) -\delta_{ij} \phi_{,0},
\quad  
\ue4{\chris{0}{0}{i}} =\psi_{,i},
\quad 
\ue5{\chris{0}{0}{0}} = \psi_{,0}+ \zeta_i \phi_{,i} .
\eeqa 

We also need
\beqa
\ue4{g_{ki,ki}} - \ue4{g_{kk,ii}} &=& 4\kappa \ue2{T^{00}} + 2 \phi_{,i}^2 - 8\(\phi\phi_{,i}\)_{,i},
\label{gij4}
\eeqa
where summations over $i$ and $k$ should be understood.
A derivation of \eqref{gij4} is given in appendix~\ref{app:PN}.


\section{Energies for massive particles}
\label{sec:res_energy}
In this section, we calculate energies for the system of massive particles in the previous section
at the 1PN, the first order in $(v/c)^2$.
We here compare three different definitions, $E_M$, $E_{\rm pt}$ and $\tilde E_{\rm pt}$ in Sect.~\ref{sec:energy}.
 

\subsection{Matter energy}
At the order up to 1PN, the matter energy is evaluated as 
\beqa
       E_M(t) &=& -\int d^3 x \sqrt{-g} T^0{}_{0} \xt \simeq
        \int d^3x \(1 + \frac{1}{2}\ue2g\) (-1 + \ue2{g _{00}})\(\ue0{T^{00}} + \ue2{T^{00}}\) \nn 
       &\simeq& \sum_{n=1}^{N} \(m_n + \frac{1}{2}m_n  \vvn^2 + m_n \phi\xnt   \), 
       \label{matter_1PN}
\eeqa
which is not conserved as
\begin{equation}
       \frac{dE_M(t)}{dt}= \sum_{n=1}^{N}m_nv_n^i \[ \partial_i \phi\xnt + \frac{d v_n^i}{dt}   \] + \sum_{n=1}^{N}m_n\partial_t \phi\xnt = \sum_{n=1}^{N}m_n\partial_t \phi\xnt,
\end{equation}
where we use \eqref{geodesic2} at 1PN as
\beqa
{d v_n^i\over dt}  = - \partial_i \phi\xnt , \quad n=1,2,\cdots, N.
\label{eom_1PN}
\eeqa
The matter energy $E_M(t)$ at this order except the rest masses $\sum_n m_n$  is nothing but the Newtonian energy of particles in the presence of the time dependent external field $\phi\xt$, and the non-conservation of this energy is understood as a violation of time translational invariance
due to the time dependent external field.  
By definition the Newtonian potential energy for the $n$-th particle is evaluated from the force as
\beqa
V\xnt &:=& -\int^{{\bf x}_n}_{\vert {\bf x}\vert=\infty} dx^i\, F_i\xt =  m_n  \int^{{\bf x}_n}_{\vert {\bf x}\vert=\infty}dx^i \, \partial_i \phi\xt
= m_n \phi\xnt,
\eeqa 
where we assume $\phi\xt \to 0$ and $V\xt\to 0$ as $\vert{\bf x}\vert\to\infty$. 
Adding the kinetic energy and summing up over $n$, the total Newtonian energy becomes
\beqa
E_{\rm Newton} &=& \sum_n  m_n\(\frac{1}{2}\vvn^2 + \phi\xnt   \),
\eeqa
which agrees with  \eqref{matter_1PN} except the rest mass.

For a conservation of a total energy,  we need other contributions absent in the matter energy,
which is probably an energy of the gravitational field. 
In the next sections, we will evaluate energies from pseudo-tensors at 1PN, in order to see how contributions from the gravitational field
are included. 

\subsubsection{A system with 2 particles}
Let us consider a system with 2 particles as an concrete example.
In this case, EOM of particles at the 1PN reduce to
\beqa
{d^2 R^i(t)\over dt^2} &=& 0, \quad 
m_R {d^2 r^i(t)\over dt^2} = - Gm_1 m_2 {r^i(t)\over r^{3}}(t),
\eeqa
where $R^i$, $r^i$ and $m_R$ are a center of mass coordinate, a relative coordinate and a reduce mass, respectively,
defined by
\beqa
R^i:={m_1 x_1^i + m_2 x_2^i\over m_1+m_2}, \quad r^i:= x_1^i -x_2^i, \quad m_R :={m_1 m_2 \over m_1+m_2}.
\eeqa 
The first equation is solved trivially as $R^i(t) = V_0^i\, t + R_0^i$, where $R_0^i$ and $V_0^i$ are a position at $t=0$ and an initial velocity, respectively.

Without loss of generality, we can take $\vr(t) = r(t)(\cos\theta(t),\sin\theta(t),0)$  to solve the second equation,
whose solution is given by
\beqa
r(t) &=&{r_0 \over 1 -e\cos\theta(t)}, \quad r_0 := {L^2\over G m_1 m_2 m_R}, \ e:=\sqrt{1+\bar E},\ \bar E:= {E\over U_0}, \
U_0:={L^2\over 2m_R r_0^2},~~~
\eeqa
where $E$ and $L$ are conserved energy and angular momentum, respectively, which are defined by
\beqa
E:= {m_R\over 2} \dot \vr^2 -{G m_1 m_2\over r(t)}, \quad L := m_R r^2(t) \dot\theta(t).
\eeqa

The corresponding matter energy is given by
\beqa
E_M(t) &=& m_1+m_2 +{m_1+ m_2\over 2}\vV_0^2+{m_R\over 2} \dot \vr^2(t) -{2G m_1 m_2\over r(t)}\nn 
&=&  m_1+m_2 +{m_1+ m_2\over 2}\vV_0^2 +E - {G m_1 m_2\over r(t)}
\eeqa

The time dependent part of $E_M(t)$, given by
\beqa
\Delta E_M(t) := - {G m_1 m_2\over r(t)} =-{ G m_1 m_2\over r_0} (1-e\cos\theta),
\eeqa
as well as its time derivative,  
\beqa
{d \Delta E_M(t)\over dt} = {Gm_1 m_2\over r^2(t)}\dot r(t) = -  {G(m_1+ m_2)Le^2\over r^2(t) r_0}\sin\theta(t),
\eeqa
are periodic in $\theta(t)$. Thus, although the matter energy is not conserved, an average of $\Delta E_M(t)$ over one period of time $T$,
which satisfies $\theta(t+T)=\theta(t)+2\pi$, vanishes.


\subsection{EH pseudo-tensor energy $E_{\rm pt}(t)$}
We first calculate the EH pseudo-tensor energy $E_{\rm pt}(t)$, which is equivalent to the generalized Komar integral for $\xi^\mu= \tau^\mu$. 
We write
\beqa
E_{\rm pt}(t) &=& E_M(t) + \Delta E(t),
\eeqa
where
\beqa
\Delta E(t) &=& -\int d^3x\, \sqrt{-g}\, t^0{}_0 =-{1\over 4\kappa}\int d^3x\, \sqrt{-g}\left[ R + g^{\alpha 0}\Gamma^\beta_{\beta 0,\alpha}
- g^{\alpha \beta}\Gamma^0_{\beta\alpha,0} \right] .
\eeqa

At the 1PN, we have
\beqa
-{1\over 4\kappa}\int d^3x\, \sqrt{-g} R &=& {1\over 2} \int d^3x\, \sqrt{-g} T \simeq -{1\over 2}\sum_n\left[ 1+\phi\xnt -{1\over 2} \vvn^2\right], 
\eeqa
where $T$ is a trace of the EMT, which satisfies $R= -2\kappa T$, and 
\beqa
&&-{1\over 4\kappa}\int d^3x\, \sqrt{-g} \(g^{\alpha 0}\Gamma^\beta_{\beta 0,\alpha}
- g^{\alpha \beta}\Gamma^0_{\beta\alpha,0} \) \simeq -{1\over 4\kappa} \int \(6 \phi_{,00} +\zeta_{i,i0}\)\nn
&=&{1\over 8\pi} \partial_t^2 \sum_n \int d^3x {m_n\over \vert  \vx -\vxn(t)\vert}=
-\sum_n {m_n\over 6} \(\vxn\cdot \dot{\vv}_n+ \vvn^2\),
\label{integral}
\eeqa
where the last integral is evaluated in appendix~\ref{app:Integration}.
Therefore we obtain 
\begin{equation}
       \Delta E(t) = \sum_n \frac{m_n}{2} \[ \frac{\vvn^2}{2}-1 -  \phi\xnt  -\frac{1}{3}\( \vxn \cdot \dot{\vv}_n + \vvn^2 \) \]
\end{equation}
at the 1PN, 
which leads to
\begin{equation}
       E_{\rm pt}(t) = \sum_n \frac{m_n}{2} \[1 + \phi\xnt  + \frac{7}{6}\vvn^2(t)  -\frac{\vxn \cdot \dot{\vv}_n}{3}  \].
       \label{pt_total}
\end{equation}

Somewhat surprisingly, it is now easily seen that $E_{\rm pt}(t)$ is {\it NOT} conserved at the 1PN as
\beqa
       \frac{dE_{\rm pt}(t)}{dt} &=& \sum_n \frac{m_n}{2} \[ \phi_{,0}\xnt + \vvn\cdot \dot{\vv}_n - \frac{1}{3}\vxn \cdot \ddot{\vv}_n  \]\nn
                             &=& -\frac{G}{6} \sum_{n,k,n \neq k} m_n m_k \frac{(\vxn - \vxk)\cdot (\vv_n-\vv_k)}{\abs \vxn - \vxk ^3}, 
\eeqa
where we use \eqref{phi-zeta}, \eqref{eom_1PN}, and their derivatives.

\subsubsection{A mechanism of the $E_{\rm pt}(t)$ non-conservation}
Since $E_{\rm pt}$ is defined as a volume integral of the conserved current, we explain a reason why it is not conserved.

For this purpose, we use a relation that  $J^\mu[\tau] \underset{\rm EOM}{=} -(T^\mu{}_0 +t^\mu{}_0)$.
We integrate the conservation equation $\partial_\mu (\sqrt{-g} J^\mu[\tau] ) = 0$
on a spacetime region $M$, whose boundary consist of two constant $x^0$ surfaces at $x^0=t_1$ and $x^0=t_2$ plus one time-like surface $M_s$ at $r:=\vert \vx\vert=\infty$, and apply the Gauss' theorem as
\beqa
0&=& \int_M d^4x\,\partial_\mu \sqrt{-g} J^\mu[\tau] = E_{\rm pt}(t_1) - E_{\rm pt}(t_2) +\int_{M_s} [d^3x]_r J^r[\tau].
\label{total_conservation}
\eeqa
If the last term remain non-zero even at $r\to \infty$, the EH pseudo-tensor energy is not conserved: $E_{\rm pt}(t_2)\not= E_{\rm pt}(t_1)$.
Using the definition of $J^\mu$ in \eqref{def_J} and  Stokes' theorem, we obtain 
\beqa
\int_{M_s} [d^3x]_r J^r[\tau] = \frac{1}{16\pi G} \int [dx]_{r} \nabla_\nu (\nabla^{[r} \tau^{\nu]}) &=& \frac{1}{16\pi G} \int_{r\to\infty} [dx]_{r0}  (\nabla^{[r} \tau^{0]}) \Big \vert ^{t_2}_{t_1}\nn 
&=& E_{\rm pt}(t_2) - E_{\rm pt}(t_1),
\label{boundary_contribution}
\eeqa 
which indeed satisfies \eqref{total_conservation},
where we use a quasi-local expression of $E_{\rm pt}(t)$ as
\beqa
E_{\rm pt}(t) &:=&{1\over 16\pi G}\int_{t} [d^3x]_0 \nabla_\nu \left(\nabla^{[0}\tau^{\nu]}\right)={1\over 16\pi G}\int_{r\to\infty,t} [d^2x]_{0r}  \left(\nabla^{[0}\tau^{r]}\right).
\eeqa
We have indeed confirmed that  an explicit evaluation of the last integral leads to \eqref{pt_total}.

Eqs.~\eqref{total_conservation} and \eqref{boundary_contribution} explain why $E_{\rm pt}$ is not conserved even though the corresponding current
is locally conserved: some contributions to $E_{\rm pt}(t)$ flow out or flow in through the spatial boundary   at $r=\infty$, so that
$E_{\rm pt}(t)$  becomes time dependent while total conservation \eqref{total_conservation} is always satisfied.
We thus finally conclude that $E_{\rm pt}(t)$ is not a good definition of the total energy for massive particles in general relativity.

\subsection{Einstein's pseudo-tensor energy $\tilde E_{\rm pt}(t)$ }
We now consider Einstein's pseudo-tensor energy $\tilde E_{\rm pt}(t)$, which is free from the second derivations of the metric.
We write
\beqa
\tilde E_{\rm pt}(t) &=& E_M(t) +\Delta \tilde E(t),\quad \Delta \tilde E(t) := -\int d^3x\, \sqrt{-g}\, \tilde t^0{}_0.
\eeqa

At the 1PN, we have
\beqa
\Delta \tilde E(t) &=&-\int d^3x\, \sqrt{-g}\, \tilde t^0{}_0 \simeq {1\over 2\kappa}\int d^3x\, \phi_{,i}^2
\label{energy_gf} \\
&=& {1\over 2\kappa} \int d^3x\,\left[ ( \phi  \phi_{,i})_{,i}  - \phi \nabla^2 \phi \right]
= -\sum_n {m_n\over 2} \phi\xnt,
\eeqa
where the total derivative term vanishes as
\beqa
\int d^3x\, \partial_i ( \phi \partial_i \phi) = \lim_{r\to\infty} \int r^2 d\Omega  {x^i\over r} \phi \phi_{,i}
= -\kappa G \lim_{r\to\infty}\sum_n{m_n\over r} =0,
\eeqa 
and we use the Einstein equation at the 1PN as
\beqa
\nabla^2\phi\xt = 4\pi G\,  \ue0{T^{00}} = 4\pi G \sum_n m_n\delta^{(3)}(\vx-\vxn).
\eeqa

Thus we obtain at the 1PN
\beqa
  \label{energy_ps2}
      \tilde E_{\rm pt}(t) &=&  \sum_n m_n \( 1 + \frac{1}{2}\phi\xnt + \frac{1}{2}\vvn^2  \),
\eeqa
where a correct factor $1/2$ appears in front of $\phi\xnt$. 
Indeed it is conserved as
\beqa
{d  \tilde E_{\rm pt}(t) \over d t} &=& {1\over 2}\sum_n \(  \phi_{,0}\xnt - v_n^i  \phi_{,i}\xnt \)
=\sum_{n\not=k}m_n m_k\( {(\vxn -\vx_k)\cdot(\vv_k +\vvn)\over \vert \vxn -\vx_k\vert^3}\),~~~~~
\eeqa
which vanishes after a summation over $n$ and $k$, since it  is anti-symmetric under  an exchange of $n$ and $k$.  

We now conclude that Einstein's pseudo-tensor energy $\tilde E_{\rm pt}(t)$ correctly reproduces the conserved energy at the 1PN
for massive point particles in general relativity, which agrees with the total energy of massive particles interacting with each other through the Newtonian potential. 
For general $b$, we have
\beqa
E_{\rm pt}^b(x^0) &:=& -\int_{x^0} [d^3x]_\mu \(\frac{1}{2\kappa} G^\mu{}_0 +t^\mu{}_0[b]\) = (1-b)  E_{\rm pt}(x^0) + b \tilde E_{\rm pt}(x^0),
\eeqa
which is not conserved unless $b=1$.
Although the current is conserved for all $b$, 
a conservation of the total energy $E_{\rm pt}^b(x^0)$ tells us that $b=1$ is a correct choice at the 1PN.
We hope that this good property holds for $\tilde E_{\rm pt}$ at higher orders.  
A factor $\dfrac{1}{2}$   of $\phi\xnt$ in  \eqref{energy_ps2}, necessary for the conservation, comes from the positive contribution
of $\Delta \tilde E(t)$, which may be interpreted as an energy of the gravitational field, as seen from  \eqref{energy_gf}.
Note however that this interpretation and the desired property of $\tilde E_{\rm pt}$ are coordinate dependent  since  $\tilde E_{\rm pt}$
is not covariant. Indeed it does not give a correct energy in a polar coordinate: It diverges even at the leading order of the PN expansion.


\subsubsection{ADM energy}
In the asymptotically flat spacetime, the total energy can be calculated using the ADM energy\cite{Arnowitt:1962hi}, defined by
\begin{equation}
  \label{ADMenergy}
  \ADM E(x^0) := \frac{1}{4\kappa} \int _{r\to \infty} [d^2 x]_{0k} \( h_{kj,j} - h_{jj,k} \)
  =  \frac{1}{4\kappa} \int  [d^3 x]_0 \( h_{kj,jk} - h_{jj,kk}\), 
\end{equation}
where $h_{\mu\nu} := g_{\mu\nu} - \eta _{\mu\nu}$ with the flat Minkowski metric $\eta_{\mu\nu}$.

At the 1PN, we have
\beqa
h_{kj,jk}-h_{jj,kk} =g_{kj,jk}-g_{jj,kk} \simeq 4 \nabla^2\phi +4\kappa  \ue2{T^{00}} + 2 \phi_{,k}^2 - 8 \(\phi  \phi_{,k}\)_{,k}, 
\eeqa
which leads to 
\beqa
 \ADM E(x^0) &\simeq& \sum_n m_n\( 1+\phi\xnt +{1\over 2} \vvn^2\) + {1\over 2\kappa} \int d^3x\,   \phi_{,k}^2 
 =\tilde E_{\rm pt}(x^0).
\eeqa
The ADM energy is conserved and agrees with $\tilde E_{\rm pt}(x^0)$ at the 1PN.

An agreement between the ADM energy and Einstein's pseudo-tensor energy can be understood as follows. 
We rewrite $\tilde E_{\rm pt}(x^0)$ as
\beqa
\tilde E_{\rm pt}(x^0) &=& {1\over 4\kappa}\int [d^3x]_0\, \partial_k\rho^k= {1\over 4\kappa} \int [d^2]_{0k} \rho^k
\eeqa
where 
\beqa
       \rho^k &= & \rg \Big [ -g^{k\mu}\chris \nu\mu\nu + \frac{1}{2}g^{\mu\nu}\chris k\mu\nu - g^{0\mu}\chris k0\mu + g^{k\mu}\chris0\mu0 
        -\frac{1}{2} \(g^{kj}\chris\nu j \nu  - g^{k\nu} \chris j\nu j - g^{j\nu}\chris k \nu j  \) \notag\\
       & -&\frac{1}{2} \( g^{k0}\chris \nu 0 \nu  - g^{k\nu} \chris 0\nu 0  - g^{0\nu} \chris k\nu 0 \) \Big ],
\eeqa
Thus we obatin
\beqa
\tilde E_{\rm pt}(x^0) &=& {1\over 4\kappa}\int  [d^2]_{0k}  \(h_{kj,j} - h_{jj,k} \) =  \ADM E(x^0),
\eeqa
showing an equivalence between the ADM energy and $\tilde E_{\rm pt}(x^0)$ beyond the 1PN,
where we estimate $\rho^k$ in the week field approximation $g_{\mu\nu} \simeq \eta _{\mu\nu} + h_{\mu\nu}$ as
 \begin{equation}
       \rho^k \simeq \(h_{kj,j} - h_{jj,k} \) + O(h^2).
 \end{equation}
 The weak field approximation gives an exact result for $\tilde E_{\rm pt}(x^0)$ at the all order of the PN expansion,
 since $h \simeq O(r^{-1})$ at large $r$ in the asymptotically flat spacetime, so that
 contributions from non-linear terms in $h$ vanishes in $\tilde E_{\rm pt}(x^0)$ as $r\to\infty$
 \footnote{Since $\rho^k$ is linear in $\chris\mu\nu\alpha$, the leading non-linear contribution has a form that $h_{**} h_{**,\beta}$,
 which is $O(r^{-3})$ at best. Note that the time derivative is $O(v r^{-1})$ in the PN expansion. 
 }.
 We thus conclude that the ADM energy is nothing but Einstein's pseudo-tensor energy
 in the asymptotically flat spacetime.
 (See also Ref.~\cite{Vollick:2021wcx}.)


\section{An extra conserved charge} 
\label{sec:res_charge}
We now turn our attention to an extra  conserved charge in  Sec.~\ref{sec:extra}, which seems different from energies in the previous section.
In this section the post-Newtonian expansion is not necessary, and therefore discussions here can be applied to systems with $\Lambda\not=0$
in an arbitrary dimension.
  
 As mentioned in Sec.~\ref{sec:extra},  it is necessary to fix the direction of $\beta^\mu$.
We here adopt the prescription proposed in \cite{Aoki:2022ysm} that $\beta^\mu$ is taken to be parallel to an energy flow of the EMT.
Explicitly we take  
\begin{equation}
       \beta^\mu (x) \simeq \[ \beta_n(s_n) + \gamma_{n,\nu}(s_n) \( x^\nu - x_n^\nu (s_n)  \)  \]v_n^\mu (s_n) + \mathcal{O}(\(x- x_n\)^2)
\end{equation}
at $x^\mu\simeq x_n^\mu(s_n)$, which gives 
\begin{equation}
      K^\mu (x) = - \sum_{n=1}^{N} \frac{m_n}{\sqrt{-g(x)}}\int ds_n\,  \beta_n (s_n) v_n^\mu(s_n) \delta^{(4)}\( x -x_n (s_n)  \).
\end{equation}
Thus the conservation of $K^\mu$ leads to 
\beqa
   \sqrt{-g} \nabla_\mu K^\mu (x) &=& \partial_\mu \{\sqrt{-g}   K^\mu (x)  \}
       = -\sum_{n=1}^{N} m_n \int ds_n\, \beta_n(s_n) v_n^\mu(s_n)\partial_\mu  \delta ^{(4)} \(x-x_n (s_n) \)  \nn
       &=& \sum_{n=1}^{N} m_n \int ds_n\, \beta_n(s_n) {d\over d s_n} \delta ^{(4)} \(x-x_n (s_n) \)\nn
       &=&  -\sum_{n=1}^{N} m_n \int ds_n\, {d \beta_n(s_n)\over d s_n}  \delta ^{(4)} \(x-x_n (s_n) \) 
       =0, 
\eeqa
which can be solved by 
\begin{equation}
       \beta_n (s_n) = -\beta_n^0 \;\;\;\; \text{for all $n$},
\end{equation}
where $\beta_n^0$ is an arbitrary real constant and a minus sign is just a convention.
 
The conserved charge is calculated as 
\begin{equation}
      Q(x^0) = \int d^3 x\, \sqrt{-g} K^0 (x) = \sum_n \beta_n^0 \int ds_n  v_n^0(s_n) \delta\(x^0-x^0_n(s_n) \)
      =\sum_{n=1}^N m_n \beta_n^0,
      \label{charge_Q}
\end{equation}
where $v_n^0 > 0$ in our definition of  the proper time.
As apparent from the expression, this charge is indeed $x^0$ independent.
Therefore there are $N$ independent conserved charges for $N$ massive particles in general relativity.  
Since $\beta_n$ has an inverse of the mass dimension in previous studies\cite{Aoki:2020nzm, Aoki:2022gez,Aoki:2022tek}, 
we here take  $\beta_n^0 =  \dfrac{1}{m_n}$ as a natural choice.
Then the conserved charge becomes a total number of particles as
\begin{equation}
 Q=\sum_{n=1}^{N} 1 =N.
 \label{number}
\end{equation}

Although the result of this extra conserved charge in  \eqref{number} looks  trivial for non-interacting massive particles,
it is non-trivial to obtain this seemingly trivial result  \eqref{number} from a generic construction of 
an extra conserved charge from the EMT in general relativity. 
We call this extra conserved charge ``a gravitational charge", as an analogy of the electric charge in electromagnetism.
It is interesting to evaluate the gravitational charge for more complicated matters with non-gravitational interactions.



\section{Conclusion}
In this paper, we have calculated the matter energy as well as the EH pseudo-tensor and Einstein's pseudo-tensor energies for the system
with massive particles in general relativity, and 
 have found that only the Einstein's pseudo-tensor energies is conserved at the 1PN.
 Non-conservations of the EH pseudo-tensor energy and others with $b\not=1$ are explained by existences of energy flows through the spatial boundary at $r=\infty$.
 It is important to note that we have also established an equivalence between  Einstein's pseudo-tensor energies and the ADM energy at all orders
 in the PN expansion.
 
 In addition, we have calculated the conserved charge for the same system without using the post-Newtonian expansion,
 and found that there are $N$ conserved quantities depending on the initial conditions. As the charge becomes the total number of particles for a natural choice of the initial condition, we call it the gravitational charge, like an electric charge in electromagnetism.
It is rather non-trivial that our generic construction of the gravitational charge from the EMT in general relativity\cite{Aoki:2020nzm} leads to 
the trivial but correct conserved charge for the system with massive particles.     

It is important to note that there always exist two conserved quantities, the energy and the gravitational charge, in general relativity.
A coexistence of two conserved quantities restrict dynamics in general relativity.
For example, while some part of matter energy can be converted to the gravitational energy such as the one for the gravitational wave,
matters never disappear completely due to the conservation of gravitational charges.
A creation of matters from gravitational fields is also prohibited.    
See Ref.~\cite{Aoki:2022ysm} for concrete examples.  

Let us discuss possible extensions of our studies in this paper.
We have established that Einstein's pseudo-tensor energy (equivalent to the ADM energy) is conserved at the 1PN in the asymptotically flat spacetime.
It is important to see a conservation of Einstein's pseudo-tensor energy at higher orders.
Since a local conservation of the corresponding current does not help, we need different  arguments to establish the conservation.
A property that Einstein's pseudo-tensor is free from second derivatives of the metric may give a hint.
Another extension is to find a conserved pseudo-tensor energy for asymptotically non-flat spacetimes such as dS or AdS.
It is interesting to calculate the gravitational charge for a system with free massless particles 
in order to see its physical meaning.  It is likely that the gravitational charge becomes a number of massless particle.
We also expect that the gravitational charge becomes more non-trivial than a number of particles for a system with interacting massive particles.
We leave these interesting and important problems to future studies. 

\section*{Acknowledgment}
SA would like to thank Prof.~Matthias Blau for useful discussions and valuable comments.
This work is supported in part by the Grant-in-Aid of the Japanese Ministry of Education, Sciences and Technology, Sports and Culture (MEXT) for Scientific Research (Nos.~JP22H00129, JP23K03387).


\appendix

\section{Derivation of \eqref{gij4}}
\label{app:PN}
We here derive $\ue4{g_{ki,ki}} - \ue4{g_{kk,ii}}$, which is necessary for the ADM energy at the 1PN
but is missing in Weinberg's textbook\cite{Weinberg:1972kfs}.

We have first obtained the following combination
 \beqa
 {1\over 2}\( 2\ue4{g_{ki,k}} - \ue4{g_{kk,i}}\) &=&\ue4{\chris ikk} -2\phi \phi_{,i} =\psi_{,i} +\zeta_{i,0} -2\phi \phi_{,i},
 \label{gkiki}
 \eeqa
using the coordinate condition  $\ue4 {g^{\alpha\beta}\chris i\alpha\beta} = 0$, which imples
 $ \ue4 {\chris ikk} = \ue 4 {\chris i00} - 4 \phi \phi_{,i} = \psi_{,i} + \zeta_{i,0}$.
 Taking one more partial derivative, we obtain
 \beqa
 {1\over 2}\( 2\ue4{g_{ki,ki}} - \ue4{g_{kk,ii}}\) &=& \kappa\(\ue2{T^{00}}+\ue2{T^{ii}}\) -3\phi_{,00}-2\(\phi \phi_{,i}\)_{,i},
 \eeqa
where we use the coordinate condition at lower order as $ \zeta_{i,i} = -4 \phi_{,0}$.

We next calculate $\ue4{g_{kk,ii}}$.
The relevant Einstein equation reads
\begin{equation}
  \ue 4R_{ij} = 2\kappa \( \ue 2 T_{ij} - \frac{1}{2}\ue2{g_{ij} T}  \),
  \label{EE4}
\end{equation}
where 
\begin{equation}
  \ue2{g_{ij} T} = \eta_{ij} \( \ue2T_{kk}  - \ue2T_{00} + 4 \phi \ue0T_{00}  \).
\end{equation}
By contracting both sides of \eqref{EE4} with $\eta^{ij}$, we obtain
\beqa
 \ue4{\chris ki{i,k}} + \ue3{\chris 0i{i,0}} - \ue4{\chris 0i{0,i}} - \ue4{\chris ki{k,i}} + \ue2{\chris kii}\ue2{\chris \alpha k \alpha} - \ue2{\chris 0i0}\ue2{\chris 0i0} - \ue2{\chris kil} \ue2{\chris lik} =  \kappa\(3\ue2{T^{00}} - \ue2{T^{kk}}\),
\eeqa
where 
\beqa
  \ue4 {\chris ki{k,i}} &=&= \frac{1}{2}  \ue4{g_{kk,ii}} - 6\(\phi\phi_{,i}  \)_{,i},\quad
  \ue3 {\chris 0i{i,0} } = \phi _{,00},\quad
  \ue4 {\chris 0i{0,i}} = \phi_{,00} + \kappa \( \ue2 {T^{00}} + \ue2 {T^{ii}}  \), \nn
  \ue2 {\chris kii} \ue2 {\chris \alpha k \alpha} &=& -2 \phi_i^2 ,\quad
  \ue2 {\chris 0i0} \ue2 {\chris 0i0} = \phi_i^2, \quad
  \ue2 {\chris kil} \ue2 {\chris lik} = - \phi_i^2.
\eeqa
We thus obtain 
\begin{equation}
  \label{4gkkii}
  \frac{1}{2}  \ue4{g_{kk.ii}} = \kappa \( \ue2 {T^{kk}} - 3 \ue2 {T^{00}}  \) + 6 \(\phi\phi_{,i} \)_{i} - 3 \phi_{,00} - 2  \phi_{,i}^2.
\end{equation}

By combining \eqref{gkiki} and \eqref{4gkkii}, we finally obtain
\begin{equation}
  \ue4 g_{ki,ki} - \ue4 g_{kk,ii} = 4\kappa \ue2{T^{00}} + 2 \phi_{,i}^2 - 8\(\phi\phi_{,i}\)_{,i}.
\end{equation}

\section{Evaluation of \eqref{integral}}
\label{app:Integration}
In this appendix we first calculate
\beqa
\partial_t \int d^3x\, {1\over \vert \vx -\vxn(t)\vert}&=& 2\pi \int r^2 dr X(r), \quad 
X(r):={1\over 2\pi} \int d\Omega {(\vx-\vxn)\cdot\vvn\over \vert \vx -\vxn\vert^3}.
\eeqa
After integrating the solid angle $\Omega$, we have
\beqa
X(r)&=& \left( {r\over r_n} I^1_3 -I^0_3\) \vxn\cdot\vvn = -{2\vxn\cdot\vvn \over r_n^3} \theta (r_n - r)
\eeqa 
where $r_n:=\vert \vxn\vert$, and 
\beqa
{1\over 2\pi}\int d\Omega{1\over \vert \vx -\vxn\vert^3}&:=&I^0_3,\quad
\quad {1\over 2\pi} \int d\Omega{x^i\over \vert \vx -\vxn\vert^3}:= {x_n^i\over r_n}  r I^1_3,
\eeqa
and
\beqa
I^0_3 &=&\left\{
\begin{array}{cc}
\dfrac{2}{r_n(r_n^2-r^2)}, &\quad r \le r_n ,\\
\\
\dfrac{2}{r(r^2-r_n^2)}, &\quad  r > r_n, \\
\end{array}
\right.
\quad
I^1_3 =\left\{
\begin{array}{cc}
\dfrac{2r}{r^2_n(r_n^2-r^2)}, &\quad r \le r_n ,\\
\\
\dfrac{2r_n}{r^2(r^2-r_n^2)}, &\quad r > r_n ,\\
\end{array}
\right. .
\eeqa

 The $r$-integral leads to
\beqa
\int dr \, r^2 X(r) = -{2\over 3}  \vxn\cdot\vvn ,
\eeqa
and we finally obtain
\beqa
\partial^2_t \int d^3x\, {1\over \vert \vx -\vxn(t)\vert}&=& -{4\pi\over 3}\partial_t \(\vxn\cdot\vvn\) =  -{4\pi\over 3}
\( \vxn\cdot\dot\vv_n +\vvn^2 \).
\eeqa




\bibliographystyle{utphys}
\bibliography{refer}

\end{document}